\documentclass[aps,amsmath,amssymb, notitlepage, 
twocolumn,
nofootinbib,
]{revtex4-1}

\usepackage[pdftex]{graphicx}
\usepackage{dcolumn}
\usepackage{bm}
\usepackage{epsfig}
\usepackage{latexsym}
\usepackage{amsmath}
\usepackage{amsfonts}
\usepackage{amssymb}
\usepackage{color}
\usepackage{array}
\usepackage{framed}
\usepackage{esvect}
\usepackage{slashed}

\newcommand{\bit}{\begin{itemize}}
\newcommand{\eit}{\end{itemize}}

\newcommand{\ba}{\begin{align}}
\newcommand{\ea}{\end{align}}
\newcommand{\be}{\begin{equation}}
\newcommand{\ee}{\end{equation}}
\newcommand{\bi}{\begin{itemize}}
\newcommand{\ei}{\end{itemize}}

\newcommand{\LL}{\mathcal{L}}

\newcommand{\T}{\mathcal{T}}

\newcommand{\nn}{\\ \nonumber}
\newcommand{\bea}{\begin{eqnarray}}
\newcommand{\eea}{\end{eqnarray}}
\newcommand{\p}{\partial}
\newcommand{\lp}{\left(}
\newcommand{\rp}{\right)}
\newcommand{\tps}{\psi_d}
\newcommand{\A}{\mathcal{A}}

\begin{document}

\title{Dual Dirac liquid on the surface of the electron topological insulator}

\author{Chong Wang and T. Senthil}
\affiliation{Department of Physics, Massachusetts Institute of Technology, Cambridge, MA 02139, USA}
\date{\today}

\begin{abstract}
We discuss a non-fermi liquid  gapless metallic surface state of the topological band insulator. It has an odd number of gapless Dirac fermions coupled to a non-compact $U(1)$ gauge field. 
This can be viewed as a vortex dual to the conventional Dirac fermion surface state. This surface duality is a reflection of a bulk dual description discussed recently for the gauged topological insulator. All the other known surface states can be conveniently accessed from the dual Dirac liquid, including the surface quantum hall state, the Fu-Kane superconductor, the gapped symmetric topological order and the `composite Dirac liquid'. We also discuss the physical properties of the dual Dirac liquid, and its connection to the half-filled Landau level.

\end{abstract}

\maketitle

 In the last decade the study of topological insulators has revealed extremely rich new physics\cite{TIs}. One of the  most interesting  is their nontrivial surface states. In the limit of non-interacting fermions, the boundary of a topological insulator is necessarily gapless as long as the relevant symmetries are unbroken. 
The classic example is the Fu-Kane-Mele (FKM) topological insulator in three dimensions\cite{FKM}, for which the two-dimensional surface can be described by a single two-component massless Dirac fermion (omitting the chemical potential term for simplicity):
\begin{equation}
\label{freedirac}
 \mathcal{L}[\psi]=\bar{\psi}(i\slashed{\partial}+\slashed{A})\psi,
\end{equation}
where $\bar{\psi}=i\psi^{\dagger}\gamma_0$, $\gamma_0=\sigma_2$, $\gamma_1=-\sigma_3$, $\gamma_2=\sigma_1$ and $\T\psi\T^{-1}=i\sigma_2\psi$. Here $A_{\mu}$ is the external probe gauge field with the $\psi$ fermions carrying physical charge $Q_{phys}=1$. 
It is well known that as long as time-reversal symmetry $\T$ and charge conservation $U(1)$ are unbroken, there is no fermion bilinear term that can gap out the $\psi$ Dirac fermions: the Dirac mass term $im\bar{\psi}\psi$ breaks time-reversal, and the pairing term $\Delta\psi^T(i\sigma_2)\psi+H.c.$ breaks charge $U(1)$ conservation. In other words, in the free fermion theory, the surface theory described by Eq.~\eqref{freedirac} is guaranteed to be gapless as long as the $U(1)\rtimes\T$ symmetries\footnote{Here the group structure $U(1)\rtimes\T$ simply means that the $U(1)$ charge is even under time-reversal.} are preserved.

%It is very interesting to ask how electron-electron interactions could change the above picture. For the theory in Eq.~\eqref{freedirac}, interacting terms are irrelevant at week coupling strength based on simple dimensional reasoning. So the real question is what could happen with strong interactions. 

The symmetry-breaking states on the surface of the FKM topological insulator are also highly nontrivial. If time-reversal is broken through a Dirac mass generation $im\bar{\psi}\psi$, where $m$ is the $\T$-breaking order parameter, the resulting state is gapped, but has a half-integer quantum hall conductance $\sigma_{xy}={\rm{sgn}}(m)\frac{1}{2}e^2/h$. This half-integer hall conductance will manifest itself on the domain wall between opposite $\T$-breaking regions, as a chiral fermion traveling in one direction on the domain wall. If time-reversal is kept but charge $U(1)$ conservation is broken instead, through a singlet pairing term $\Delta\psi^T(i\sigma_2)\psi+H.c.$, the resulting surface becomes the famous Fu-Kane superconductor\cite{FuKane}, which is gapped, with vortices trapping Majorana zero modes.

Both the gapless Dirac surface and the symmetry breaking surfaces are adequately understood within (almost) free fermion theory. In recent years, the effect of strong interactions on the surface has attracted much attention. It was realized that under strong interactions, the surface of many topological insulators can be gapped while preserving all the relevant symmetries. The price we pay by symmetrically gapping out the surface is to introduce intrinsic topological order, i.e. with fractionalized anyon excitations on the surface. This was first demonstrated in bosonic analogues of the topological insulators\cite{avts12,hmodl,burnellbc,chenanomaloussymm} that are also known as symmetry-protected topological phases\cite{chencoho2011,atav13,sptannrev}, and subsequently in fermionic systems\cite{fidkowski3d} including the FKM topological insulator\cite{fSTO1,fSTO2,fSTO3,fSTO4}. More recently, another interesting interacting surface state called `composite Dirac liquid' was proposed\cite{cdl}. This is a 
gapless surface state with gapped charge excitations, and various topologically ordered states are obtained upon further gapping out the theory.

The common theme behind all the different surface states is the parity anomaly\cite{parityanomaly} first discussed in the field theory literature in the context of a single Dirac cone. Due to the anomaly, all those nontrivial surface states, with their symmetry properties, cannot be realized in strictly two dimensional systems without the presence of the bulk topological insulator. 

In this paper we propose another interacting surface state of the FKM topological insulator, which we call the `dual Dirac liquid'. It is a non-fermi liquid metallic state that allows us to conveniently access all the other surface states described above within a unified framework. It has an extremely simple field theory description, with one (or more generally an odd number)  Dirac cone coupled to a non-compact $U(1)$ gauge field $a_{\mu}$:
\begin{equation}
\label{ddl}
 \mathcal{L}_{Dual}[\tps,a_{\mu}]=\bar{\tps}(i\slashed{\partial}+2\slashed{a})\tps+\mathcal{L}_{Maxwell}[a_{\mu}]+\frac{\epsilon^{ijk}}{2\pi}A_i\p_ja_k.
\end{equation}
where $\bar{\tps}=i\tps^{\dagger}\gamma_0$, $\gamma_0=\sigma_2$, $\gamma_1=-\sigma_3$, $\gamma_2=\sigma_1$ and $\T\tps\T^{-1}=e^{i\phi}\sigma_2\tps^{\dagger}$. Notice that the time-reversal action on the dual $\tps$ fermions is very different from the original fermions $\psi$: the $U(1)$ gauge charge of $\tps$ is odd under time-reversal ($\T:q_{dual}\to-q_{dual}$), in contrast to the physical charge carried by $\psi$, which is even under $\T$. Also, time-reversal action on $\tps$ has a $U(1)$ gauge ambiguity $e^{i\phi}$, which comes from the $U(1)$ gauge symmetry in the theory. Since the gauge charge is $\T$-odd, the $U(1)$ gauge rotation commutes with time-reversal. In particular, this implies that $\T^2=e^{i2\phi}$, which is gauge non-invariant and hence physically meaningless. It is easy to check that such a time-reversal implementation, together with the $U(1)$ gauge symmetry, prevents any fermion bilinear mass term in the theory. Another crucial point of the theory is that the minimal charge under the 
$a_{\mu}$ gauge field is $q_{min}
=2$ and is carried by the $\tps$ fermion. In particular, dual charge-$1$ object does not exist in the spectrum of the dual Dirac liquid.

Eq.~\eqref{ddl} can be viewed as a theory of vortices\cite{dual}, where $2\pi$-flux of the $a_{\mu}$ gauge field carries physical charge $Q_{phys}=1$ of the original electric charge, while the $\tps$ fermions behave as four-fold ($4\pi$ or $2hc/e$) vortices. The fermionic statistics of the four-fold vortex can be understood by thinking about a monopole of magnetic charge $Q_m=2$ in the bulk, since the vortex can be created on the surface by tunneling a monopole across the surface. It is well known\cite{qi} that the topological insulator induces a $\theta$-term in the bulk electromagnetic field at $\theta=\pi$, which in turn gives a polarization charge to the monopole $Q_{e}=\frac{\theta}{2\pi}Q_m=Q_m/2$ through the Witten effect\cite{witten}. For $Q_m=2$ the polarization charge is $Q_e=1$, which can be canceled by attaching an electron to it. The resulting charge-neutral $Q_m=2$ monopole is therefore a fermion due to the fermion statistics of the electron\cite{maxswe,fSTO1}.

Below we will describe more details of the dual Dirac liquid. We show that all the known surface states can be accessed from the dual Dirac liquid, including the surface quantum hall state, the Fu-Kane superconductor, the gapped symmetric topological orders and the composite Dirac liquid. This also serves as a `proof' of the existence of the dual Dirac liquid as a legitimate surface state, since one can always imagine reversing those phase transitions. We then discuss the understanding from the bulk perspective in more detail, and discuss several subtle issues at the end.

\section{Accessing the known surface states}

\subsection{Quantum hall magnet} 
\label{qhe}

Consider breaking time-reversal symmetry by introducing a Dirac mass term in Eq.~\eqref{ddl}: $\Delta\mathcal{L}=m\tps^{\dagger}\sigma_2\tps$. The $\tps$ fermions are now gapped. A gapped single Dirac fermion will induce a Chern-Simons term for the gauge field $a_{\mu}$, which is known in the field theory literature as parity anomaly\cite{parityanomaly}. The level of the Chern-Simons term here is $k=\frac{1}{2}q_{dual}^2{\rm{sgn}}(m)=2{\rm{sgn}}(m)$ because the $\tps$ fermion carries dual charge $q_{dual}=2$:

\be
\label{CS}
\LL_{CS}={\rm{sgn}}(m)\frac{2}{4\pi}\epsilon^{ijk}a_i\p_ja_k.
\ee

Such a Chern-Simons term will also gap out the gauge field $a_{\mu}$, so the entire surface state is now gapped. So what are the excitations of this gapped state? Naively one may think there are fractionalized anyons in the excitation spectrum since there is a Chern-Simons theory at level larger than one. However, one should remember that in Eq.~\eqref{ddl} the minimum charge under $a_{\mu}$ is $q_{min}=2$ and is carried by a fermion $\tps$. So the minimum excitation is created by the $\tps$ fermion. The Chern-Simons term in Eq.~\eqref{CS} will bind a flux of $2\pi$ to the fermion $\tps$. But since $\tps$ carries $q_{dual}=2$ under $a_{\mu}$, this flux-binding will not change its statistics. However, from Eq.~\eqref{ddl} we see that a $2\pi$-flux of the $a_{\mu}$ field means physical charge $Q_{phys}=1$ in the probe $A_{\mu}$ gauge field. Therefore this fermionic excitation is nothing but the original $\psi$ fermion (the electron)! Since there is no other nontrivial excitation, the surface is just 
the familiar non-fractionalized magnetic insulator on the TI surface. This can also be seen from the surface Hall conductance: integrating out the $a_{\mu}$ field in the presence of the Chern-Simons term in Eq.~\eqref{CS} gives a Chern-Simons term for the probe gauge field $A_{\mu}$ at level $-\frac{1}{2}{\rm{sgn}}(m)$, which implies a Hall conductance $\sigma_{xy}=-\frac{1}{2}{\rm{sgn}}(m)$ -- the well known result for the surface magnet.

\subsection{Fu-Kane superconductor}

Since Eq.~\eqref{ddl} looks very similar to the boson-vortex duality\cite{dual}, the natural question is how to access a superconductor from the dual Dirac liquid. In the ordinary dual-vortex theory, when the vortex is gapped while keeping the $a_{\mu}$ gauge field gapless, the system is in a superfluid state, with the $a_{\mu}$ photon being the Goldstone mode. If we try to apply this idea to the dual Dirac liquid, we immediately encounter an issue: there is no fermion bilinear term that can gap out the $\tps$ fermions while keeping the time-reversal and $U(1)$ gauge symmetry. However, it was realized recently\cite{fidkowski3d,3dfSPT2,maxvortex} that the fermions, under strong interaction, can develop a gap while preserving the $U(1)\times\T$ symmetry\footnote{The group structure $U(1)\times\T$ means that the $U(1)$ gauge charge is odd under $\T$.}. The resulting gapped state will have intrinsic topological order, with anyons in the excitation spectrum. While there are more 
than one 
possible 
surface topological orders, it turns out to be useful to discuss the topological order proposed in 
Ref.~\cite{fidkowski3d} which we call the dual T-Pfaffian state. %But we will call it the dual T-Pfaffian since it is obtained through gapping out the dual Dirac cone, and reserve the name T-Pfaffian for the state obtained by gapping out the original Dirac cone.

When the $\tps$ fermions are gapped, we obtain a superconductor on the surface. The question now is what kind of superconductor we have obtained. Naively since the gapped $\tps$ fermions form a fractionalized topological order, one might expect the resulting superconductor to also have fractionalized topological order. However, if all the fractionalized quasi-particles are either vortices (charged under $a_{\mu}$), or the Bogoliubov fermion $f$ that has a mutual $\pi$-statistics with the $\pi$-vortices (particles carrying dual charge $q_{dual}=1/2$), the resulting superconductor will be BSC-like and not fractionalized\cite{z2long}. It turns out that the dual T-Pfaffian state belongs to this case: all the fractional quasi-particles in T-Pfaffian are charged under $a_{\mu}$, except a fermion $\epsilon$ that has $\pi$-statistics with particles carrying dual charge $q_{dual}=1/2$.\footnote{This would not be true if we gapped out the $\tps$ fermions with another surface topological order obtained from vortex-
condensation\cite{3dfSPT2,
maxvortex}, which has 
charge-neutral fractional quasi-particles beyond the Bogoliubov fermion $f$.} We show below that the resulting state is indeed the famous Fu-Kane superconductor obtained from pair-condensing the Dirac cone in Eq.~\eqref{freedirac}.

%The details of the dual T-Pfaffian state is reviewed in Appendix~\ref{dtpfaffian}. 

The dual T-Pfaffian state has several key features. First, there is a unique minimally charged particle $v$ (under $a_{\mu}$) with charge $q_{dual}=q_{min}/4=1/2$. This particle is an Ising-Majorana-like anyon, in the sense that
\bea
\label{fusion1}
v\times\epsilon&\sim&v, \nn
v\times v&\sim& s\times(1+\epsilon),
\eea
where $\epsilon$ is a charge $q_{dual}=0$ fermion, $s$ is a charge $q_{dual}=1$ semion, and $s\times\epsilon$ is an anti-semion. The first fusion rule indicates that the $\epsilon$ fermion has a mutual $\pi$-statistics with the $v$ particle, and there is a Majorana zero-mode trapped in the $v$ particle. An important fusion rule for the semion $s$ is
\be
\label{fusion2}
s\times s\times\epsilon\sim\tps.
\ee
This implies that four copies of $s$ ($s^4$) is a trivial charge $q_{dual}=4$ boson, in the sense that it has no nontrivial braiding statistics with anything. It can also be shown\cite{fidkowski3d} that the $\epsilon$ fermion must have $\T^2=-1$ to be consistent with the fusion rules.

In the unit of vorticity, charge $q_{dual}=1$ particles correspond to $2\pi$-vortices. Therefore the $v$ particle is a $\pi$-vortex -- exactly what one expects for a BCS superconductor. The $\pi$-mutual statistics between $\epsilon$ and $v$ means that the $\epsilon$ particle should be interpreted as the Bogoliubov fermion, which is consistent with the Kramers degeneracy for $\epsilon$. 

The facts that (1) the $\pi$-vortex $v$ is an Ising-Majorana-like anyon, (2) the $2\pi$-vortices $s$ and $\epsilon s$ are semions and anti-semions, respectively, and (3) the eight-fold ($8\pi$) vortex $s^4$ is topologically trivial, match exactly with the vortex properties\footnote{Strictly speaking, the quantum statistics are only formally assigned to the field operators corresponding to the vortices. The real vortex, which is coupled to the gapless $a_{\mu}$ field (the charge fluctuation), does not have sharply defined abelian statistics.} of the Fu-Kane superconductor, as discussed in detail in Ref.~\cite{fSTO1,fSTO2}. Therefore the seemingly complicated superconductor we obtained is nothing but the Fu-Kane superconductor obtained from pair-condensing the free Dirac cone!

%More importantly, this vortex $v$ is an Ising-Majorana-like anyon, in the sense that fusing two such anyons produces either a $2\pi$-vortex ($q=1$ under $a_{\mu}$), or a $2\pi$-vortex bound with a neutral fermion $\epsilon$. So this particle $v$ is exactly the Majorana vortex in the Fu-Kane superconductor!

%One can check that other details of the gauged dual T-Pfaffian state matches with the Fu-Kane superconductor perfectly. In fact, the Fu-Kane superconductor, in the vortex dual description developed in Ref.~\cite{fSTO1,fSTO2}, is exactly the gauged dual T-Pfaffian state. We review this in Appendix~\ref{fukane}.

\subsection{T-Pfaffian topological order}

 Now consider keeping time-reversal symmetry, but pair-condensing the $\tps$ fermions with

\be
\label{pairing}
\Delta\LL=\Delta\tps^T(i\sigma_2)\tps+H.c.,
\ee
which gaps out both the $\tps$ fermions and the $a_{\mu}$ gauge field, and forces time-reversal to act as $\T\tps\T^{-1}=\sigma_2\tps^{\dagger}$, which has $\T^2=1$ on the fermions. 
The fermions now form a `dual' Fu-Kane superconductor, in which the vortex properties are almost the same with the familiar Fu-Kane superconductor, with fusion rules identical to that described in Eq.~\eqref{fusion1}. But time-reversal acts differently: since the $U(1)$ gauge charge changes sign under $\T$, the vorticity is unchanged under $\T$. In particular, the fundamental Ising-Majorana vortex is time-reversal invariant: $\T v\T^{-1}=v$. 
The two-fold vortex $s$ is a semion, which under time-reversal becomes another two-fold vortex $s\tps$ which is an anti-semion.
%More details are reviewed in Appendix~\ref{dualfukane}. 

Since the `dual' Fu-Kane superconductor is coupled with the gauge field $a_{\mu}$, the vortices are gapped and short-range interacting quasi-particles. Therefore the gapped surface has intrinsic topological order, and the anyons are simply the vortices and the Bogoliubov $\tps$ fermion. 
Since the condensate carries charge $q_{dual}=2q_{min}=4$ under $a_{\mu}$, the minimum vortex carries flux $\pi/2$ under $a_{\mu}$. From Eq.~\eqref{ddl} we see that $\pi/2$-flux of $a_{\mu}$ means physical charge $Q_{phys}=1/4$ under the probe gauge field $A_{\mu}$. Therefore the minimally charged quasi-particle $v$ is an Ising-Majorana-like anyon that carries physical charge $Q_{phys}=1/4$ and is time-reversal invariant. The two-fold vortices $s$ and $s\tps$ becomes $Q_{phys}=1/2$ semion and antisemion, and switch to each other under time-reversal. The combination $s^2\tps$ is a $Q_{phys}=1$ fermion that has no nontrivial braiding statistics with anything, therefore should be interpreted as the physical fermion $\psi$. 
These properties match perfectly with that of the T-Pfaffian surface topological order\cite{fSTO3}. Therefore we have obtained the T-Pfaffian state from the dual Dirac liquid through pair-condensing the dual fermions. 
%We review the details of the T-Pfaffian state in Appendix~\ref{tpfaffian}. By comparing the details, it is straightforward to see that the gauged dual Fu-Kane superconductor produces exactly the T-Pfaffian state. 
In particular, the somewhat mysterious neutral fermion with $\T^2=1$ in the T-Pfaffian state is rationalized from this approach: $\T^2=1$ is necessary for $\tps$ to keep the pairing term Eq.~\eqref{pairing} invariant.

It is interesting to view the above discussions as an exotic form of the superconductor-insulator duality: the superconductor of the $\tps$ fermions corresponds to a topologically ordered insulating state, while the topologically ordered insulator of the $\tps$ fermions corresponds to a superconductor.

\subsection{Composite Dirac liquid}

 Now consider pair-condensing the $\tps$ fermions without gapping them out. This can be achieved by having a paring amplitude non-zero at UV scale but vanishes near the Dirac point. For example, we can have
\be
\label{uvpairing}
\Delta\LL=\Delta\tps^T(i\partial^3_x)\tps+H.c.,
\ee
which gaps out the $a_{\mu}$ gauge field but not the $\tps$ fermions, anf forces time-reversal to act as $\T\tps\T^{-1}=\sigma_2\tps^{\dagger}$ with $\T^2=1$. 
The pairing term Eq.~\eqref{uvpairing} is irrelevant at the Dirac point, so the low energy theory is simply the dual Dirac fermion in Eq.~\eqref{ddl}, with the gauge field $a_{\mu}$ gapped. The gaplessness of the dual Dirac fermion in this case is not protected, but nevertheless it serves as a parent state for other surface phases.

We now show that the UV paired dual Dirac liquid in Eq.~\eqref{uvpairing} is exactly what has been discussed recently in Ref.~\cite{cdl} as the `composite Dirac liquid'. The composite Dirac liquid has three key features: (1) the charge degrees of freedom are gapped even though the state itself is gapless, (2) gapping out the fermions by pairing produces the T-Pfaffian nonabelian state, and (3) gapping out the fermions by a $\T$-breaking Dirac mass term gives an Abelian $(113)$ hierarchical state. Now we show that the paired dual Dirac liquid indeed reproduces all these features.

First, since the $a_{\mu}$ field describes charge fluctuation and is gapped due to the pair condensate at UV scale, the charge degrees of freedom are obviously gapped in this phase. If we further gap out the dual Dirac fermion by a pairing term, we simply reproduce the physics obtained from Eq.~\eqref{pairing}, where we showed explicitly that the resulting state is just the T-Pfaffian topological order. So we are left with only one thing to check: the $\T$-breaking Dirac mass phase. 

The $\tps$ fermions in this phase form a weakly paired $\T$-breaking superconductor, so the fundamental excitations are the Bogoliubov fermion $\tps$ and the vortex $v$ of the pair condensate. The vortices are short-range interacting due to the coupling to the $a_{\mu}$ field. Since the condensate charries charge $q_{dual}=2q_{min}=4$ under $a_{\mu}$, the fundamental vortex $v$ carries flux-$\pi/2$ of $a_{\mu}$, which means physical charge $Q_{phys}=1/4$ under the original electron charge. 
Now what about the statistics of $v$? The gapped Dirac cone produces a Chern-Simons term for the $a_{\mu}$ field at level $k={\rm{sgn}}(m)\frac{1}{2}q_{min}^2=2$, as discussed in Eq.~\eqref{CS}. This means that the $\pi/2$ flux will carry self statistical (exchange) phase $\theta_v={\rm{sgn}}(m)\frac{\pi}{8}$. So the final theory has a charge-$1/4$ anyon $v$ with exchange phase $\theta_v={\rm{sgn}}(m)\frac{\pi}{8}$, a charge-neutral fermion $\tps$, with a mutual $\pi$-braiding phase between the two since $\tps$ is now a Bogoliubov fermion. This is exactly the $(113)$-hierarchical state (or its conjugate depending on ${\rm{sgn}}(m)$), described by the $K$-matrix and charge vector $\tau$
\be
K=\pm\lp\begin{array}{cc}
          1 & 3 \\
          3 & 1 \end{array}\rp, \hspace{5pt} \tau=\lp\begin{array}{c}
                                                        1 \\
                                                        1 \end{array}\rp.
\ee
The $\pm\frac{\pi}{8}$ anyon $v$ is labeled by the particle vector $l^T=(1,0)$ and the Bogoliubov fermion is labeled by $l^T=(1,-1)$. This is in full agreement with that obtained in Ref.~\cite{cdl}.

\section{Relation to the bulk}

We now discuss the bulk correspondence in more detail. We start by coupling the bulk topological insulator to a compact $U(1)$ gauge field $\A_{\mu}$. As discussed in the earlier part of this paper, the pure monopole with magnetic charge $Q_m=2$ becomes a fermion due to Witten effect\cite{maxswe,fSTO1}. Traditionally this bulk system -- often known as the topological Mott insulator\cite{pesinlb} -- is understood as a $U(1)$ gauge theory in which the fermionic charge form a topological insulator. However, it was realized recently\cite{wstoappear} that this very same state can also be viewed as a `dual' topological insulator of the fermionic monopole. Essentially, this is because of an electric-magnetic duality in the $U(1)$ gauge theory: both the $Q_e=1$ charge and the $Q_m=2$ monopole are fermions, so we can interchange the two, then rescale the charge unit by $1/2$ (hence rescale the magnetic charge unit by $2$), and the resulting theory will be identical to the original one in terms 
of charge-monopole spectrum.

Once we view the state as a dual topological insulator of the $Q_m=2$ monopole, the surface theory can immediately be deduced from the surface state of the dual topological insulator, which is simply one Dirac cone $\tps$ with time-reversal implemented as $\T\tps\T^{-1}=i\sigma_2\tps^{\dagger}$. The total surface state is simply this Dirac cone coupled with the bulk gauge field, which is almost identical to that in Eq.~\eqref{ddl}, except for the difference that the gauge field is only on the surface but not in the bulk.

%It may be possible to make the correspondence between the bulk electro-magnetic duality and the surface dual Dirac liquid more precise, by thinking holographically, as suggested in Ref.~\cite{witten03}. But we shall not go deep into that direction.

 \section{Physical properties} 
 
Notice that the same physics discussed so far can be reproduced if we modify Eq.~\eqref{ddl} to contain any odd number of Dirac fermions coupled to the gauge field $a_{\mu}$. It is known that if the flavor number $N_{\tps}=2n+1$ is large enough, the dual Dirac liquid is described by a stable conformal field theory at low energy. It should be understood as an algebraic vortex liquid\cite{avl1,avl2,avl3,avlhermele,avlws} and will have many of the same physical properties. It will have a finite universal electrical conductivity in the zero temperature limit, and the compressibility will vanish linearly with the temperature. The physical electron operator should be understood as an instanton operator at this fixed point, and will have a non-trivial scaling dimension. The electron Green's function will therefore have a non-fermi liquid form.  Thus the surface state we have described at large $N_{\tps}$ is a non-fermi liquid metal with anomalous implementation of the physical symmetries.
 
The physics at smaller $N_{\tps}$, especially when $N_{\tps}=1$, is less clear. One may worry about the stability of the dual Dirac liquid in this case, since the gauge coupling is expected to be relevant in Eq.~\eqref{ddl}. However, notice that the usual mechanisms for instability (chiral symmetry breaking and confinement) may both be absent here: there is no chiral symmetry at N = 1 to break, and the theory is non-compact so confinement is also not an option. Therefore it may be possible for the phase to be stable, even at $N_{\tps}=1$.

For $N_{\tps}=1$, there is yet another interesting possibility: the dual Dirac liquid at $N_{\tps}=1$ may simply be the free Dirac fermions, namely Eq.~\eqref{ddl} and Eq.~\eqref{freedirac} may actually describe the same phase! This option is made more plausible by the result in Sec.~\ref{qhe}: both phases can be constructed by proliferating domain walls between quantum Hall magnets with $\sigma_{xy}=\pm1/2$. 

The two possibilities for $N_{\tps}=1$ are both very interesting. The first option leads to a new phase, and the second one leads to a new representation of the familiar free Dirac phase, which then leads to new phase transitions like the one to the T-Pfaffian state. At this point we are not able to decide which of these options for $N_{\tps}=1$ is actually realized.

\section{Discussion}

\subsection{Subtlety of $N_{\tps}$}

A subtle issue is that there is actually an ambiguity in the bulk state that supports the dual Dirac liquid in Eq.~\eqref{ddl}: it is either the FKM topological insulator, or the combination of the FKM topological insulator with a bosonic symmetry-protected topological state known as the $eTmT$ topological paramagnet\cite{avts12,hmodl}. Equivalently, according to Ref.~\cite{3dfSPT2,maxvortex}, there is an ambiguity on the suface of the FKM topological insulator: the number of fermion flavors in the dual Dirac liquid should either be $N_{\tps}=8n\pm1$ or $N_{\tps}=8n\pm3$. The two possibilities corresponds to different bulk, but we could not decide which one is the bare FKM insulator. This issue is related to the ambiguity of time-reversal action in the T-Pfaffian state on the FKM surface\cite{fSTO3}.

\subsection{Relation to half-filled Landau level}

With a very different motivation and perspective, the same Lagrangian in Eq.~\eqref{ddl} was proposed by Son in Ref.~\cite{son} as a particle-hole symmetric formulation of the half-filled Landau level. This symmetry is not manifest in the  classic theory of the half-filled Landau level\cite{hlr}. It was also realized in Ref.~\cite{son} that pair-condensing the dual Dirac fermions leads to a non-Abelian topological order dubbed PH-Pfaffian. We now briefly discuss how the present work is connected to Son's proposal. A more elaborate discussion of the half-filled Landau level will be presented in a separate work\cite{wsll}.

Consider Eq.~\eqref{freedirac} in the presence of another anti-unitary symmetry called particle-hole symmetry $C$ in Eq.~\eqref{freedirac}, where
\be
C\psi C^{-1} =\sigma_2 \psi^{\dagger},
\ee
This will describe the surface (within band theory) of a bulk topological insulator with symmetry $U(1) \times C$.  Note that with this symmetry the chemical potential is {\em required }to be zero. 

Now with this symmetry external $B$-fields are $C$-symmetric and hence can be included in the surface action.  The surface Dirac fermions will then form Landau levels, including famously one which is exactly at zero energy.  Now the $C$-symmetry guarantees that this zeroth Landau level is exactly half-filled. At low energies we can project to this Landau level.  The degeneracy of many body states in the half-filled Landau level will be resolved by interactions. Thus the low energy physics is exactly the classic problem of the half-filled Landau level. The key point is that the half-filled Landau level when thus obtained has emerged in a system which microscopically has the particle-hole $C$ symmetry. This should be contrasted with conventional realizations of the half-filled Landau level where particle-hole is an emergent low energy symmetry of a single Landau level. 
The fate of the half-filled Landau level is thus tied to the fate of correlated surface states of the {\em three dimensional} topological insulator with $U(1) \times C$ symmetry. 

The results in this paper provide an understanding of Son's proposal. First return to Eq.~\eqref{freedirac} in the presence of the $C$ symmetry but without any external $B$-field. Note that the action of $C$ is similar to how time-reversal is implemented on the dual Dirac fermions. The dual Dirac liquid in Eq.~\eqref{ddl} will still be a legitimate surface state, and the particle-hole symmetry will act on $\tps$ in the same way as time-reversal acts on $\psi$:
\be
\label{phdd}
C \tps C^{-1} =i\sigma_2\tps.
\ee

Now consider turning on a non-zero magnetic field in Eq.~\eqref{freedirac} to get the half-filled Landau level.  Interestingly such a magnetic field is readily incorporated in the dual Dirac liquid side: in the dual picture, the physical magnetic field is represented by a chemical potential term in Eq.~\eqref{ddl}, which gives a fermi surface, with fermi area equal to half of the total flux. Particle-hole symmetry is clearly preserved in this picture, which acts simply as Eq.~\eqref{phdd}.

\textbf{Acknowledgment}: This work was supported by NSF DMR-1305741.  This work was partially supported by a Simons Investigator award from the
Simons Foundation to Senthil Todadri. During the completion of this manuscript, we became aware of a similar work by M. Metlitski and A. Vishwanath\cite{maxashvin}, which discusses the same dual Dirac liquid state on the surface of topological insulator. We thank them for sharing their unpublished work.

%\newpage

\end{document}